\documentclass{icrc2009}

\usepackage{graphicx}   
\usepackage[caption=false]{caption}    
\usepackage[font=footnotesize]{subfig} 
\usepackage{fixltx2e}
\usepackage{url}

\newcommand{\shorttitle}[1]%
{\markboth{Proceedings of the 31\MakeLowercase{$^{st}$} ICRC, {\L}\'{o}d\'{z} 2009}{#1} }
\newcommand{\etal}{\MakeLowercase{\textit{et al. }}} 

\begin{document}
\title{Monte Carlo Study on the Large Imaging Air Cherekov
Telescopes for $>$ 10 GeV gamma ray astronomy}

\author{\IEEEauthorblockN{M.Teshima\IEEEauthorrefmark{1},
			  E.Carmona\IEEEauthorrefmark{1},
                          P.Colin\IEEEauthorrefmark{1}, 
                          P.Majumdar\IEEEauthorrefmark{2} and 
                           T.Schweizer\IEEEauthorrefmark{1}}
                            \\
\IEEEauthorblockA{\IEEEauthorrefmark{1}Max Planck Instit\"ut f\"ur   
     Physik, M\"unchen, Germany}
\IEEEauthorblockA{\IEEEauthorrefmark{2}Deutsches Elekronen-Synchrotron,
          Zeuthen, Germany }}

\shorttitle{Teshima \etal Monte Carlo Study}
\maketitle

\begin{abstract}
The Imaging Air Cherenkov Telescopes (IACTs), like, HESS, MAGIC and
VERITAS well demonstrated their performances by showing many exciting
results at very high energy gamma ray domain, mainly between 100 GeV
and 10 TeV. It is important to investigate how much we can improve the
sensitivity in this energy range, but it is also important to expand
the energy coverage and sensitivity towards new domains, the lower and
higher energies, by extending this IACT techniques. For this purpose,
we have carried out the optimization of the array of large IACTs
assuming with new technologies, advanced photodetectors, and Ultra
Fast readout system by Monte Carlo simulation, especially to obtain
the best sensitivity in the energy range between 10 GeV and 100 GeV. We
will report the performance of the array of Large IACTs with advanced
technologies and its limitation.

  \end{abstract}

 
\section{Introduction}

Recently, ground-based very high-energy (VHE) $\gamma$-ray astronomy
achieved a remarkable advancement in the development of the observational
technique for the registration and study of $\gamma$-ray emission above
100 GeV. The sourthern hemisphere HESS array of four 12~m IACTs, located in Namibia,
and the northern hemisphere VERITAS array of four similary designed telescopes,
located in Arizona, have proven many outstanding advantages of
stereoscopic observations of VHE $\gamma$ rays with the ground-based
detectors. The MAGIC telescope, consisting of a single
17~m telescope, equipped with a fast response, high-resolution imaging camera,
and located at La Palma in the Canary Islands,
has demonstrated its performance, comparable to that of HESS and VERITAS, in
observations of $\gamma$ rays with energies above 100~GeV in addition to
the unique capability of detecting $\gamma$-ray showers ranging in
energy well below 100~GeV, though at a limited sensitivity level.
Currently, the MAGIC collaboration is commissioning a second 17~m telescope on
the same site in order to enable stereoscopic observations, which will drastically
boost the sensitivity of future detections of $\gamma$-ray fluxes in the
sub-100~GeV energy domain. Based on Monte Carlo simulations, we have studied 
the response of an array of few large telescopes of larger aperture, e.g., 
20-25 $m$ class in conjunction with an array of modest medium sized telescopes of
12 $m$ class. We investigate the key parameters, i.e. sensitivity, angular resolution, 
energy resolution of such a system, primarily in the sub-100 GeV range and also try to 
improve the sensitivity at energies above 100 GeV.     
 
\section{Simulations}

The Monte Carlo
simulation of the array is divided into three stages. The {\em
CORSIKA}~\cite{corsika} program simulates the air showers initiated by
either high energy gammas or hadrons.  We have used the CORSIKA version
6.500 for our simulations, the EGS4 code for
electromagnetic shower generation and QJSJET-II and FLUKA
for high and low energy hadronic interactions respectively. We have also
used new atmospheric models for MAGIC on the basis of studies of
total mass density as a function of the height.  The second stage of the
simulation, {\em Reflector } program, accounts for the Cherenkov light
absorption and scattering in the atmosphere and then performs the
reflection of the surviving photons on the mirror dish to obtain their
location and arrival time on the camera plane. Finally, the {\em
Camera} program simulates the behaviour of the MAGIC photomultipliers,
trigger system and data acquisition electronics. Realistic pulse
shapes, noise level and gain fluctuations obtained from the MAGIC
real data have been implemented in the simulation software.

For the present study a total of 2.1 $\times$ 10$^6$ gammas between
20 GeV and 50 TeV have been produced, as well as and 4.5 $\times$
10$^7$ protons between 10 GeV and 50 TeV. The energy distribution of
both primary gamma rays and protons is a pure power law with a spectral index 
of -2.0. The telescope
pointing direction is uniform between 0 and 30$^\circ$ in zenith, 
with the directions of protons scattered isotropically within a $6^\circ$ 
semi-aperture cone
around the telescope axis. Maximum impact parameters of 1 km and 1.5~km
have been simulated for gammas and hadrons respectively. Each shower was 
reused 10 and 20 times for gammas and protons respectively using the shower reuse 
option in CORSIKA. The contribution of electrons was neglected. To this 
has been added the contribution of
Helium and heavier nuclei which is estimated to be $\sim$ 50\% of the proton rate. 

\begin{figure}[!h]
  \centering
  \includegraphics[width=2.5in]{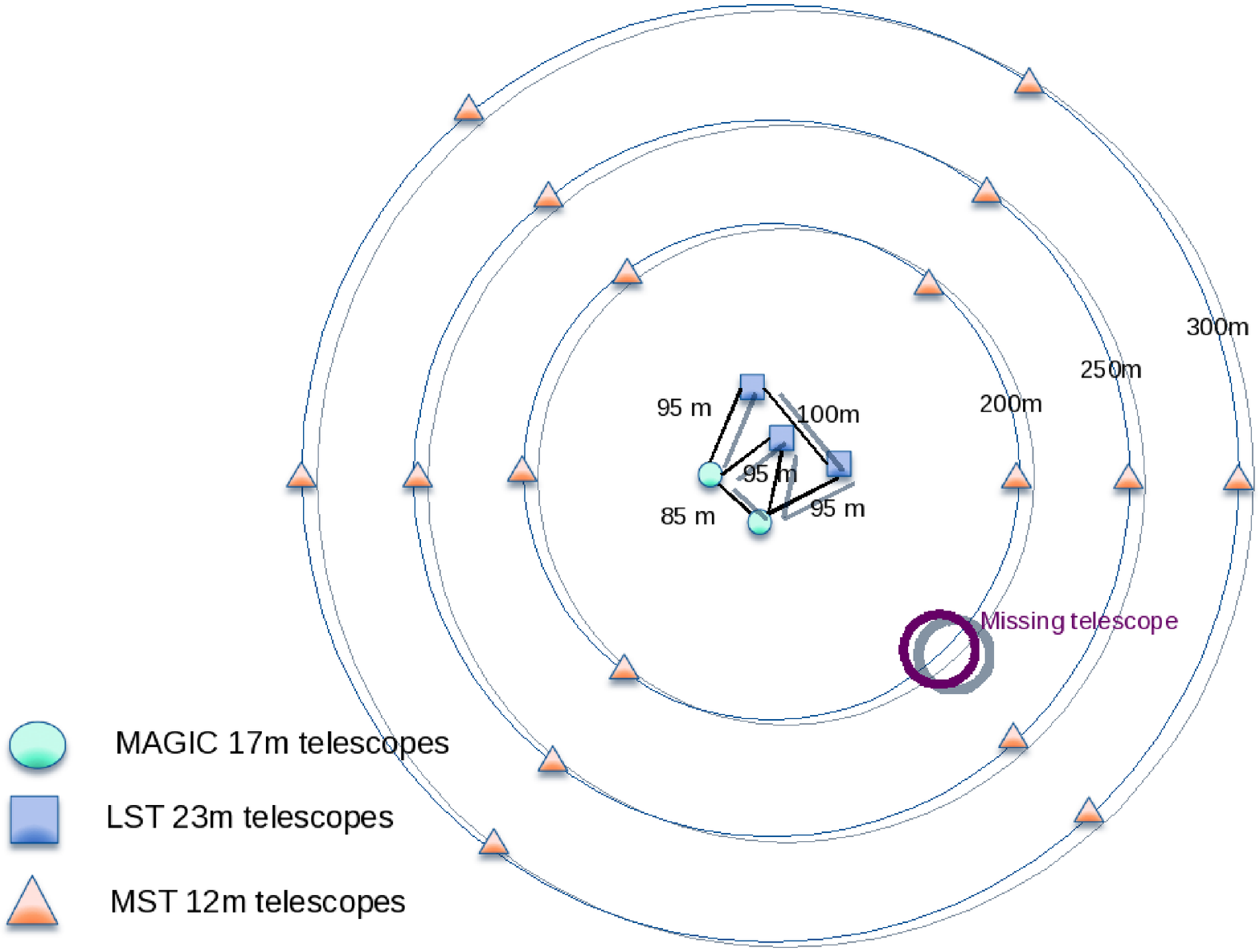}
  \caption{Simulated Array Configuration}
  \label{array}
 \end{figure}

The array configuration studied here is shown in Figure~\ref{array}. 
In addition to the two 17 m MAGIC telescopes (from now on MAGIC-2) which are separated by 
a distance of 85 m, we have considered different cases by placing a 23 m telescope 
at a distance of 85 m from the MAGIC telescopes by making a triangle configuration. 
In addition we have two 23 m telescopes Large Size Telescopes (from now on LST ) 
separated by a distance of 100 m which makes 
a trapezoidal configuration with the MAGIC telescope system. We have also placed 6 
telescopes of medium size structure ( $\sim$ 12 m class ) in a circular pattern 
at a distance of 200 m from the centre of the array. The performance of the large
telescope array system along with these small telescopes have been studied for 2 more 
distance configurations, viz., by placing the small telescopes at a distance of 250 m and 300 m 
respectively (from now on MSTRing1, MSTRing2 and MSTRing3 for distances 200, 250 and 300 m respectively).    
One telescope in the first ring had to be removed because of a problem in the assignment of the 
coordinate system for that particular telescope. 
The f/D of the MAGIC-2, LST and 12 m class telescopes are 1.0, 1.2 and 1.0 repectively. The 
FoV of the telescopes are 3.5, 5 and 5 degrees respectively and the camera is equipped 
with 1039, 2245 and 859
pixels with super bialkali high quantum efficiency photomultipliers which has a peak 
quantum efficieny of $\sim$ 32\%. The pixel sizes for the large telescopes are 0.1$^\circ$ deg. 
and that for the 12 m class are 0.2$^\circ$ deg.  

\section{Analysis of Stereo Events }

The stereoscopic observation mode allows a precise reconstruction of the shower
parameters as well as a stronger suppression of the hadronic showers and
other background events. 
The analysis of stereoscopic events is performed by individually analyzing the
images from each telescope. A set of parameters (Hillas
parameters~\cite{hillas1985}) is obtained from each image and they are
combined to obtain the shower parameters. Only showers triggering 
the large telescopes are considered under the stereo analysis. 
The trigger threshold is estimated to be $\sim$ 25 GeV when 2 LSTs participate 
in the trigger. The images from each telescope are
combined following the algorithm in~\cite{hofmann-1999-122}. The
intersection point of the major axes of the ellipses recorded in
the telescope cameras, provides the location of the source of a
particular shower. The
location of the shower core on ground is obtained by intersecting the
image axes from the telescope positions on the ground. In addition,
the height of the shower maximum ($H_{max}$) can also be
obtained.

The data analysis was carried out using the standard MAGIC
analysis and reconstruction software MARS~\cite{2009ICRCMARS}.
Before image parametrization, a
tail-cut  cleaning of the image  was performed, requiring signals
higher than a pre-defined absolute amplitude level and time
coincidences with neighbouring channels. The time
coincidence effectively suppresses pixels containing only noise 
from the night sky. The procedure used in this work can be summarized as :
\begin{itemize}
\item After selecting the core pixels, we reject those whose arrival time is not within
  a time $\Delta t_1$ of the mean arrival time of all core pixels.
\item In the selection of the boundary pixels we add the constraint
  that the time difference between the boundary pixel candidate and
  its neighbor core pixels is smaller than a second fixed time
  constraint $\Delta t_2$.
\end{itemize}
For most of the analysis the minimum required
pixel content is 6 phe for so-called core pixels and 3 phe for
boundary pixels for the MAGIC type telescopes and 12 m class telescopes.
For the 23 m telescopes, the pixel content was raised to 9 phe and 5 phe
for core and boundary pixels respectively.
The choice of these values is supported by a study based on Monte Carlo data
for the single dish MAGIC telescope~(see \cite{ICRC$_$timing} and \cite{tescaro$_$tesina} for more details).
{\em Width} and {\em Length}, the most important parameters for 
gamma/hadron separation, depend on the distance from the shower
core to the telescope. Hence a correct estimation of the impact
parameter is required to properly evaluate these parameters. With a
single telescope, the observer can not easily resolve the ambiguity
between a close by, low energy shower and a distant, high energy
one. With two or more telescopes, in most cases the ambiguity disappears
because of the stereoscopic vision of the showers.

In order to combine the parameters from both images, we compute the
{\em Mean Scaled Width}(MSW) and {\em Length}(MSL) parameters. These new
parameters are obtained by subtracting the mean and dividing by the
RMS of the parameter distribution for Monte Carlo gammas. While combining 
the images from different telescopes, a cut of 40 $phe$ was applied to the
images to throw away the very small images which may be contaminated by a lot
of noise and hence are very difficult to reconstruct. 

To reject the hadronic background a multi-tree classifier algorithm 
based on the "Random Forest (RF)" 
method~\cite{Bre01,Bock04, MAGICRF} was used for the $\gamma$/hadron separation. The 
selection conditions were trained with Monte Carlo simulated $\gamma$-ray 
samples~\cite{Maj05} and a sample
of experimental background events.
In the RF method for each event the so-called HADRONNESS parameter (H) was calculated 
from a combination of all the image parameters. H assigns to each event a number between 
0 and 1 of being more hadron-like (high H) or $\gamma$-ray like (low H values). 
The selection of events with low H value enriches $\gamma$-ray events in the surviving data sample.
For this study, the parameters that have been used in Random Forest
are: average amplitude ($Size$), core distance, $H_{max}$, $MSW$ and
$MSL$. These set of parameters, however, should not be considered as
optimal and some improvement could be expected with an optimized
parameter selection. In the final analysis, gamma/hadron separation is based on the
$Hadronness$ and $\theta^2$ parameter~\footnote{$\theta^2$
is defined as the square of the angular distance from the real source
image in the camera and the reconstructed one for each event}
is used to extract the signal events.

\section{Simulation Results}

\begin{figure}[!h]
  \centering
  \includegraphics[width=2.7in]{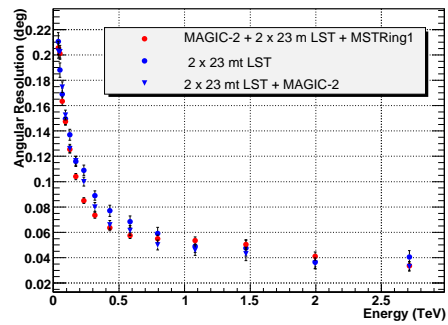}
  \caption{Angular Resolution for the Cerenkov Telescope Array for different configurations 
   of the array (see text for details)}
  \label{angres1}
 \end{figure}

The sensitivity is defined as ``integral flux resulting in gamma excess
events, in 50 hours of observation, equals to 5 times the standard
deviation of the background". While estimating the sensitivity, a conservative number 
up to 2\% systematic error has been taken into account for each flux point. 
All the sensitivity numbers are quoted in terms of percentages of Crab (standard candle in Very High 
Energy $\gamma-$ ray astronomy). While computing the sensitivities, we have restricted
our analysis up to 500 GeV as beyond that it is difficult to compute a meaningful sensitivity
due to lack of backround events after application of cuts. 
The stereoscopic technique allows for a better sensitivity below 100 GeV and also a
reduction of the analysis threshold is achieved over a single telescope~\cite{colin}.
One of the biggest advantages of stereoscopy is that the
direction of gamma rays is better reconstructed.
For a single telescope, the angular resolution is estimated
using a modified parametrisation
of the so called {\it $DISP$} method~\cite{domingo}.
However, this method suffers from a drawback
as a result of which a certain fraction of gamma events are
misreconstructed.
With two or more telescopes, this drawback is easily
overcome since the source direction is obtained as the intersection of
major axes of the images in the camera.
The distribution
of arrival directions for simulated $\gamma$-ray showers can be fitted to a
two-dimension Gaussian function. The $\sigma$-parameter of the Gaussian fit stands for the
angular resolution of the directional reconstruction. The angular resolution as a function 
of gamma ray energy is shown in Figure~\ref{angres1}. While estimating the angular resolution 
a minimum of 2 telescopes is required, the angular resolution improving substantially
when the minimum number of required telescopes is increased. From the Figure~\ref{angres1}
it is seen that the angular resolution at and below 100 GeV improves significantly 
for the configuration MAGIC-2+LST+MSTRing1 over the other configurations and also for MAGIC-2 system
alone~\cite{colin}.      

\begin{figure}[!h]
  \centering
  \includegraphics[width=2.7in]{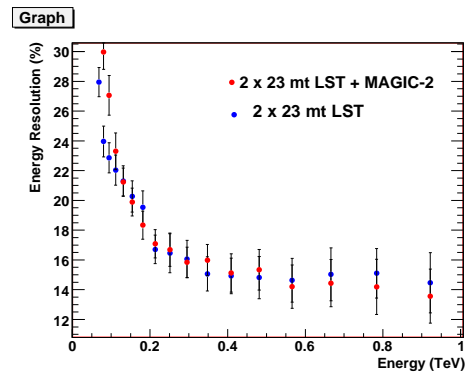}
  \caption{Energy Resolution for the Cerenkov Telescope Array for different configurations
          of the array (see text for details)}
  \label{energy}
 \end{figure}

The stereoscopic analysis also results in a better energy reconstruction
over a single telescope due to better estimate of the shower axis and also a 
multiple sampling of the light pool.
The energy is reconstructed for each telescope with lookup tables based on 
image size ( i.e. number of photoelectrons in the image ), impact parameter, 
height of the shower maximum and zenith angle. These lookup tables are built
from Monte Carlo simulations of $\gamma-$ ray showers.
The energy resolution for
gammas as a function of primary energy is shown in Figure~\ref{energy} 
for two different cases :\\

  \begin{itemize}
   \item 2 $\times$ 23 m LST 
    \item 2 $\times$ 23 m LST + MAGIC-2
  \end{itemize}
An energy resolution of $\sim$ 15\% is achieved above 500
GeV. It must be noted that both the angular resolution and the energy 
resolution improves significantly if the minimum number of telescopes  
in the analysis is raised, however, in such a case, the energy threshold of 
the primary also increases depending on the number of telescopes used. 

\begin{figure}[!h]
  \centering
  \includegraphics[width=2.7in]{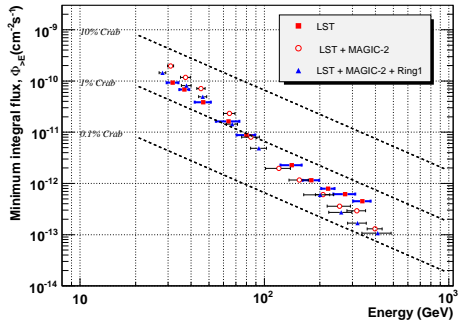}
  \caption{Sensitivity of the Cerenkov Telescope Array for different configurations
        of the array. The best sensitivity is achived for the case MAGIC-2+LST+MSTRingX(X=1,2,or 3), shown by 
         blue triangles.(see text for details) }
  \label{sensitivity}
 \end{figure}

Figure~\ref{sensitivity} shows the sensitivity of the array for the 
following configurations : \\
  \begin{itemize}
   \item  2 $\times$ 23 m LST 
    \item 2 $\times$ 23 m LST + MAGIC-2
    \item 2 $\times$ 23 m LST + MAGIC-2 + Ring1
  \end{itemize}

\begin{figure}[!h]
  \centering
  \includegraphics[width=2.7in]{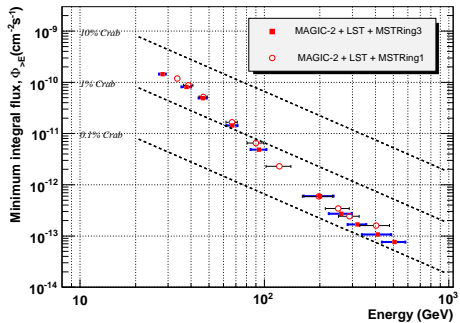}
  \caption{Sensitivity of the Cerenkov Telescope Array for different ring configurations. Both the ring configurations 
    give similar sensitivities.}
  \label{sens_ring}
 \end{figure}

It must be noted that a very standard analysis of stereoscopic data has been performed here and
a better optimisation of the analysis over different energy ranges is expected to 
improve the sensitivity of different configurations.
It is seen that the sensitivity for all three configurations improve over
a system of 2 MAGIC telescopes~\cite{colin} at 100 GeV and above. Specially above 100 GeV,
the LST+MAGIC-2+Ring1 configuration achieves the best sensitivity ($\sim$ a few milliCrab) 
for the cases studied here. Below 100 GeV, the improvement in sensitivity over a system of 
two MAGIC telescopes is very modest as compared to a system of MAGIC-2 telescopes only~\cite{colin} 
reaching to about 3-4\% of Crab at around 80 GeV.  
We have also investigated the effect of the rings, however both the telescope configurations 
with MSTRing1 and MSTRing3 yielded comparable sensitivities (see Figure~\ref{sens_ring}). 

\begin{figure}[!h]
  \centering
  \includegraphics[width=2.9in]{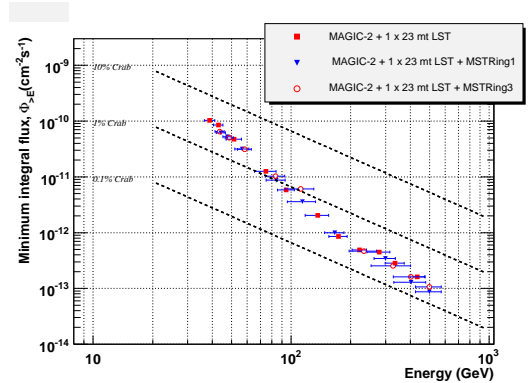}
  \caption{Sensitivity of the Cerenkov Telescope Array for different configurations
        of the array (see text for details)}
  \label{sensitivity1}
 \end{figure}

Figure~\ref{sensitivity1} compares the sensitivities of the configuration 
of MAGIC-2 + 1 $\times$ 23 m LST telescopes with 2 different rings added to the 
3 telescope configuration.  
It is seen that at higher energies ($>$ 100 GeV), 
the sensitivity improves with the addition of the ring, whereas there is marginal or no improvement 
at energies below 100 GeV. 
In addition, comparing the sensitivity curves for Figure~\ref{sensitivity} 
and Figure~\ref{sensitivity1}, one can see that the curves
follow a very similar pattern apart from the fact that one is able to reach a lower threshold in the first case 
by the addition of a large 23 mt telescope. Thus, more the number of bigger telescopes, lower the 
energy threshold one can reach.  
We have also studied the sensitivities for the array configuration of 3 $\times$ 23 mt LST 
with and without outer rings added to it. However, this configuration does not yield 
good sensitivity due to the very nature of the configuration where the third telescope is placed
very close to the other two 23 mt LST. (see Figure~ref{array})  

\section{Conclusions}
We have performed extensive Monte Carlo simulations to study the response and 
performance of an array of Large Imaging Air Cerenkov Telescopes. This IACT can serve as a 
prototype for future low energy ground based $\gamma-$ ray experiment at $>$ 10 GeV. The angular 
resolution of such a system has been shown to be 
significantly better than MAGIC-2 sytem of telescopes
at around 100 GeV. The sensitivity of the system has been shown to be better than MAGIC-2 reaching  
below 1\% of Crab around 
100 GeV and improves significantly at moderately high energies ($\sim$ few milliCrab below 800 GeV). 
So, future arrays of large           
telescopes along with few medium sized ones can be efficient enough for effective detection
of $\gamma-$ rays of a few tens of GeV and can be very competetive instruments.   
However, below 100 GeV, the improvement in sensitivity is found to be modest. A sophisticated 
analysis to improve the sensitivity below 100 GeV is required and is currently under investigation.

\end{document}